\documentstyle[aps,aps,epsf]{revtex}

\begin{document}

\newcommand\1{$\spadesuit$}
\newcommand\2{$\clubsuit$}
\tighten
\draft

\twocolumn[\hsize\textwidth\columnwidth\hsize\csname 
@twocolumnfalse\endcsname

\title{The Robustness of Quintessence}

\author{Philippe Brax}
\address{Service de Physique Th\'eorique, CEA-Saclay, F-91191 Gif/Yvette Cedex, France. \\
e-mail: brax@spht.saclay.cea.fr}

\author{J\'er\^ome Martin}
\address{DARC, Observatoire de Paris-CNRS UMR 8629, 92195 Meudon Cedex, France. \\
e-mail: martin@edelweiss.obspm.fr}

\date{July 30, 1999}
\maketitle

\begin{abstract}
Recent observations seem to suggest that our Universe is accelerating implying 
that it is dominated by a fluid whose equation of state is negative. Quintessence is 
a possible explanation. In particular, the concept of tracking solutions permits 
to adress the fine-tuning and coincidence problems. We study this proposal in 
the simplest case of an inverse power potential and investigate its robustness
to corrections. We show that quintessence is not affected by
the one-loop quantum corrections. In the supersymmetric case where the quintessential 
potential is motivated by non-perturbative effects in gauge theories, we consider the 
curvature effects and the K\"ahler corrections.
We find that the curvature effects are negligible while the K\"ahler corrections 
modify the early evolution of the quintessence field.
Finally we study the supergravity corrections and show that they must be taken 
into account as  $Q\approx  m_{\rm Pl}$ at small red-shifts. We discuss simple 
supergravity models exhibiting the quintessential behaviour. In particular, 
we propose a model where the scalar potential is given 
by $V(Q)=\frac{\Lambda^{4+\alpha }}{Q^{\alpha }}e^{\frac{\kappa}{2}Q^2}$. We argue 
that the fine-tuning problem can be overcome if $\alpha \ge 11$. This 
model leads to $\omega _Q\approx -0.82$ for $\Omega _{\rm m}\approx 0.3$ which 
is in good agreement with the presently available data.

\end{abstract}
\narrowtext
\vspace{1cm}]

\section{Introduction}

Several observations seem to suggest that our 
present Universe is dominated by a type of matter with negative 
equation of state, $\omega _Q\equiv p_Q/\rho _Q<0$. The first 
type of observations leading to this conclusion is the recent measurements 
of the relation luminous distance versus redshift using type Ia 
supernovae \cite{SNIa}. The interpretation of the data are usually 
made under the assumption that the unknown fluid is a ``true'' cosmological 
constant $\Lambda $. Unfortunately, the results are degenerate in the 
$\Omega _{\rm m}-\Omega _{\Lambda }$ plane and it is difficult to draw a
conclusion  on the basis of these measurements only. The situation changes 
drastically if one includes in the analysis a second type of observations: the 
measurements of the CMB anisotropies. In this case the degeneracy can be 
removed \cite{Teg} and one is led to the conclusion that the matter with 
negative equation of state would contribute by $70 \%$ to the total energy 
of the Universe, the remaining $30 \%$ 
being essentially Cold Dark Matter ensuring that the Universe is spatially 
flat, $\Omega _0=1$, in agreement with the standard inflationary 
scenario. This conclusion can be traced back to the fact 
that many CMB experiments show a 
high amplitude of the first Doppler peak located at $\ell \approx 260$. For 
example, this is the case for the experiments Saskatoon \cite{Sask}, 
PythonV \cite{PythonV} or TOCO97 \cite{Toco}. The addition of a fluid with negative 
equation of 
state has for consequence that the Integrated 
Sachs-Wolfe effect reinforces the scales $\ell \approx 200-300$ and increases the 
peak to values compatible with the error bars of these experiments. Moreover, the 
position of the peak informs on the value of $\Omega _0$ where $\Omega _0$ is the 
ratio of the total energy density to the critical energy density. In the simplest 
case the position is predicted to be $\ell_{\rm Doppler} 
\approx 220/\sqrt{\Omega _0}$. In the case where a fluid with negative equation of state 
is added the peak is shifted towards bigger values of $\ell $. This seems to be the 
case for the experiments cited above. 
\par
Recently, another method was proposed in order to remove the degeneracy 
between $\Omega _{\rm m}$ and $\Omega _{\Lambda }$ using large-scale 
peculiar velocities data \cite{velo}. These data provide constraints mainly 
on $\Omega _{\rm m}$ and are almost independent of $\Omega _{\Lambda }$. Combined 
with the measurement of the relation luminous distance versus redshift, they 
select a region in the $\Omega _{\rm m}-\Omega _{\Lambda }$ plan which is 
compatible with the results of Ref. \cite{Teg}. It is remarkable that, although 
of different nature, these experimental 
data converge towards the same conclusion. 
\par
This raises the issue of the physical origin of this fluid with negative 
equation of state. A useful indicator of the physical nature of this 
fluid is the value of $\omega _Q$. Recent constraints \cite{WCOS} indicate that 
$-1\le \omega _Q \le -0.6$ whereas in Refs. \cite{PTW,Eft} a value such 
that $-1\le \omega _Q \le -0.8$ is favoured. The case $\omega _Q=-1$ corresponds 
to the existence of a ``true'' non-zero cosmological constant. This cosmological 
constant has then to be explained by current particle physics 
scenarios. In particular one has to face the task of explaining an energy scale of 
$\approx 5.7h^2 \times 10^{-47} \mbox{GeV}^4$, i.e.~a value far from the natural 
scales of particle physics. Therefore, although perfectly compatible with the 
presently available data, this hypothesis runs into theoretical problems 
since it seems easier to explain a vanishing cosmological constant (by some 
yet unknown fundamental mechanism maybe coming from  quantum gravity or string theory, see 
Ref. \cite{Coleman}) than finding a reason for a tiny (in comparison with the high energy 
physics scales) contribution. In a certain sense the measurements described 
above render the ``quantum'' cosmological constant problem worse than before.
\par
Recently, another explanation, named quintessence, has been put forward 
in Refs. \cite{quint}. Quintessence is an alternative scenario with a homogeneous scalar field 
$Q$ whose equation of state is such that $-1 \le \omega _Q \le 0$. In this scenario, 
the missing energy density is due to this scalar field. Let us note that this 
explanation allows to come back to the situation where there is a vanishing cosmological 
constant. However the quintessence scenario does not solve the ``quantum'' cosmological 
constant problem. 
\par
Quintessence has to adress a certain number of questions. First of all one must 
make sure that the fine tuning problem of the cosmological constant does not reappear 
in a different guise. One must also solve the ``coincidence problem'', i.e.~understand 
why the quintessential field begins to dominate now. Another conundrum 
is to try to justify the presence of such a field from the Particle 
Physics point of view. The answers to these questions strongly depend on the form 
of the potential $V(Q)$. For example, if one chooses a potential of the form 
$V(Q)=(1/2)m^2Q^2$ then one cannot avoid to fine tune the value of the mass to 
an extremely small number \cite{KL}. The problem is then similar to the case of the 
cosmological constant.
\par
However, the problems described previously can be adressed if one considers the 
following potential \cite{track}:
\begin{equation}
\label{invpot}
V(Q)=\frac{\Lambda ^{\alpha +4}}{Q^{\alpha }},
\end{equation} 
where $\alpha \ge 0$ and $\Lambda $ are free parameters. This potential 
possesses remarkable properties. The equations of motion have an attractor 
solution called in Ref. \cite{track} the ``tracking field''. The initial conditions 
can vary by $100$ orders of magnitude leading to the attractor in all the cases. Since 
the present value of $Q \approx m _{\rm Pl}$ on the attractor, one has 
that $\Lambda \approx (\Omega _Q \rho _{\rm c}m_{\rm Pl}^{\alpha })^{1/(4+\alpha )}
\approx 10^6$GeV for $\alpha =6$ where 
$\rho _{\rm c}=8.1h^2 \times 10^{-47}\mbox{GeV}^4$ is 
the present value of the critical energy density. This value is not in contradiction with 
usual high energy scales. Moreover, one can hope to justify the form of the potential 
given in Eq. (\ref{invpot}) from high energy physics \cite{PB,us,Choi,MPR}. Finally it 
should be noted 
that, in principle, it is possible to
distinguish quintessence from a cosmological constant since one has in 
general $\omega _Q\neq -1$. 
\par
The aim of this paper is to study the robustness of the concept of tracker solutions.
In section II we quickly review the main properties of the tracking solutions. Then 
in section III, we analyze whether the nice properties of the tracking field are affected 
by the quantum corrections to the potential given by Eq. (\ref{invpot}) at the one loop 
level in the case where the underlying model is not supersymmetric. We 
show that the quintessential scenario is robust against these corrections. In section 
IV, we turn to the study of the SUSY models. We argue, as already noted in 
Ref. \cite{PB}, that 
potentials given by Eq. (\ref{invpot}) naturally 
arise in the context of supersymmetric gauge theories where certain flat directions 
are lifted by 
non-perturbative effects. We study the phenomenology of these models 
for which the quantum corrections to the superpotential 
automatically cancel out. As quintessence requires a value of the second derivative of the 
potential of 
the order of the scalar curvature of the universe, we also study the  corrections 
due to the fact that the fields live in a curved spacetime. The curvature effects are 
evaluated at 
the one-loop level and shown to preserve the tracker 
field properties. Finally in the supersymmetric case one can take into account the effect 
of 
the 
corrections to the kinetic terms of the quintessence field. In particular in the low 
energy description of the supersymmetric gauge theory the K\"ahler 
potential receives corrections suppressed by the gaugino condensation scale.  We show 
that this leads to difficulties for the supersymmetric 
models of the tracker potential. In section V, we turn to the study of SUGRA models 
of quintessence. We emphasize that such models are the most physical 
ones since at the end of the evolution the field is on tracks which implies that its 
value today is $Q\approx m_{\rm Pl}$. We analyse the 
SUGRA corrections to the inverse power law potential and show that they lead to 
inconsistencies due to the possible negative values of the potential. To remedy 
this situation we propose a supergravity scenario where the potential is guaranteed to 
remain positive. We apply this framework to the case of the heterotic string where the 
role of the quintessence field is played by the string moduli. Indeed the moduli are 
famous for leading to run-away potentials
as expected for quintessence. We find that the resulting potential is exponentially 
decreasing, a case already studied in the literature which fails to give the 
appropriate energy density. We eventually present a toy model 
where the inverse power law results from the SUGRA potential. We end with 
the conclusions presented in section VI.

\section{Tracking solutions}

In this section we quickly review the main properties of the tracking 
solutions as explained in Refs. \cite{track}. 
\par
The Universe is described by a spatially flat 
Friedmann-Lemaitre-Robertson-Walker (FLRW) spacetime whose metric 
can be written as: ${\rm d}s^2=-{\rm d}t^2+a^2(t){\rm d}{\bf x}^2$.
\par
We assume that the matter content 
of the Universe is composed of five different fluids: baryons, cold dark matter, photons, 
neutrinos and the quintessential field $Q$. The energy density of Baryons and cold 
dark matter evolves as 
$\rho _{\rm m}=\rho _{\rm c} \Omega _{\rm m}(1+z)^3$ where $z$ is the redshift. The 
equation of state is $p_{\rm m}=0$ which is equivalent to $\omega _{\rm m}=0$. Observations 
indicate that 
$\Omega _{\rm m}=\Omega _{\rm b}+\Omega _{\rm cdm}\approx 0.3$. Photons and neutrinos 
have an energy 
density given by $\rho _{\rm r}=\rho _{\rm c} \Omega _{\rm r}(1+z)^4$. The 
equation of state is given by $\omega _{\rm r}=1/3$. The contribution of radiation 
is negligible today since $\Omega _{\rm r}=\Omega _{\gamma }+\Omega _{\nu }
\approx 10^{-4}$. Finally, the fifth component is the scalar 
field $Q$. Its equation of state is characterized by $\omega _Q=
[\frac{1}{2}\dot{Q}^2-V(Q)]/[\frac{1}{2}\dot{Q}^2+V(Q)]$ where a dot represents 
a derivative with respect to the cosmic time. {\it A priori}, 
$\omega _Q$ is not a constant and is such that $-1 \le \omega _Q \le 1$. Since 
the Universe is supposed to be spatially flat, we always have $\Omega _{\rm m}
+\Omega _{\rm r}+\Omega _Q=1$ which leads to $\Omega _Q\approx 0.7$. In the following, we 
will denote 
the dominant component in the energy 
density by $\rho _{\rm B}$ so that during the radiation dominated era we have $\rho _{\rm B}=
\rho _{\rm r}$ and during the matter dominated era, $\rho _{\rm B} =\rho _{\rm m}$. A similar 
notation will be used for $\omega _{\rm B}$.
\par
The evolution of the scale factor 
is governed by the Friedmann equation:
\begin{equation}
\label{Friedmann}
H^2=\biggl(\frac{\dot{a}}{a}\biggr)^2=\frac{\kappa }{3} (\rho _m+\rho _r+\rho _Q),
\end{equation}
where $\kappa \equiv 8\pi G/c^4=8\pi /m_{\rm Pl}^2$ in the Planck system 
of units. The evolution 
of the scalar field is given by the Klein Gordon equation:
\begin{equation}
\label{KG}
\ddot{Q}+3H\dot{Q}+V'(Q)=0,
\end{equation}
where a prime denotes the derivative with respect to $Q$.  
\par
The inverse power law potential was first studied in Ref. \cite{RP}. If one 
requires that, during the radiation dominated era, the energy density 
of the scalar field be subdominant (this is necessary for not being in conflict with 
the Big Bang Nucleosynthesis), i.e. $\rho _Q \ll \rho _{\rm B}$, and redshift as 
$\rho _Q\propto a^{-4\alpha /(\alpha +2) }$ then one is automatically led to 
the potential of Eq. (\ref{invpot}). This was the original motivation of Ref. \cite{RP} 
for considering the potential (\ref{invpot}). In 
that case, it is possible to 
find an exact solution to the Klein Gordon equation for which 
$Q\propto a^{4/(\alpha +2)}$. One can show that this solution 
is an attractor \cite{RP}. Then, if one follows the behaviour of the scalar 
field during the matter dominated era (i.e. for $\rho _{\rm B}=\rho _{\rm m}$) with the same 
potential, one can show \cite{RP} that $Q \propto a^{3/(\alpha +2)}$ is an exact solution 
which is still an attractor. For this solution, one has $\rho _Q \propto a^{-3\alpha 
/(\alpha +2)}$. The previous results are equivalent to saying that the attractor is 
given by:
\begin{equation}
\label{attractor}
\frac{{\rm d}^2V(Q)}{{\rm d}Q^2}=\frac{9}{2}\frac{\alpha +1}{\alpha }(1-\omega _Q^2)H^2,
\end{equation}
during both the radiation and matter dominated epochs\footnote{ Establishing this relation 
from Ref. \cite{RP} the factor 
$1/2$ is no longer present because the definition of the potential in that paper 
differs by this factor from the definition adopted in Ref. \cite{RP} and 
in the present article}.  We can re-write the parameter 
$\omega _Q$ as $\omega _Q=(\alpha \omega _{\rm B}-2)/(\alpha +2)$. Since $\rho _Q$ 
redshifts slower than $\rho _{\rm B}$, the scalar field contribution becomes 
dominant at some stage of the evolution.
\par
As shown in Ref. \cite{track}, this scenario possesses important advantages. Firstly, 
as already stated in the introduction, one can hope to avoid any fine-tuning. Indeed 
if the scalar field is on tracks today and begins to dominate and if, in addition, we 
require $\Omega _Q\approx 0.7$ then 
 $\Lambda \approx 4.8 \times 10^{6}\mbox{GeV}$ (for $\alpha =6$), a very reasonable scale 
from the High Energy Physics point of view. Secondly, the solution will 
be on tracks today for a huge range of initial conditions. If one fixes the 
initial conditions at the end of inflation, $z=10^{28}$, the allowed initial values for the 
energy density are such that $10^{-37}\mbox{GeV}^4 \le \rho _Q \le 10^{61}
\mbox{GeV}^4$ where $10^{-37}\mbox{GeV}^4$ is approximatively the background energy density at 
equality whereas $10^{61}\mbox{GeV}^4$ represents the background energy 
density at the initial redshift. If the scalar field starts at rest, this means 
that $10^{-18}m_{\rm Pl} \le Q_{\rm i} \le 10^{-2}m_{\rm Pl}$ initially. Thirdly, the 
value of $\omega _Q$ is automatically such that $-1 \le \omega _Q \le 0$ 
today. The precise value of $\omega _Q$ depends on the functional form of $V(Q)$ 
and on the value of $\Omega _{\rm m}$.
\par
In the following figures, we illustrate these properties for $\alpha =6$, 
$\Omega _{\rm m}=0.3$ and $Q_{\rm i}(z=10^{28})\approx 3\times 10^{-18}m_{\rm Pl}$ which 
roughly corresponds to equipartion at that redshift, that is to say $\Omega _{Qi}\approx 
10^{-4}$. Equations (\ref{Friedmann}) and (\ref{KG}) are 
integrated numerically. The first figure represents the evolution of the energy 
densities throughout the radiation and matter dominated epochs.

\begin{figure}
\begin{center}
\leavevmode
\hbox{%
\epsfxsize=8cm
\epsffile{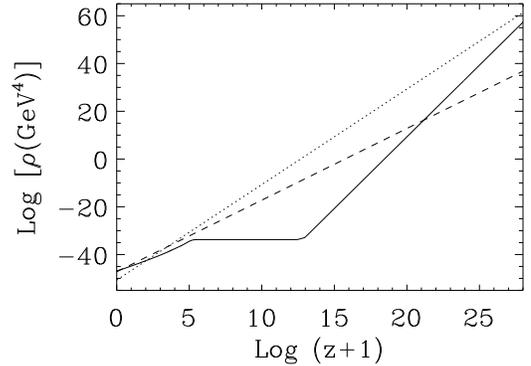}}
\end{center}
\caption{Energy density of radiation, matter and quintessence in function 
of the redshift starting from equipartition. Dotted line represents radiation, 
dashed line is matter and the full line is quintessence.}
\label{rho0,17}
\end{figure}

The interpretation of these curves has already been given in Ref. \cite{track}. The 
case presented here corresponds to an ``overshoot'' according to the terminology 
of that reference. First the scalar field rolls down the potential such that 
its kinetic energy dominates and $\dot{Q}\propto a^{-3}$. Then, the field freezes 
to some value $Q_{\rm f}$. And finally, it joins the attractor. 
\par
In the next figure, the evolution of the equation of state for the same model 
is displayed. 

\begin{figure}
\begin{center}
\leavevmode
\hbox{%
\epsfxsize=8cm
\epsffile{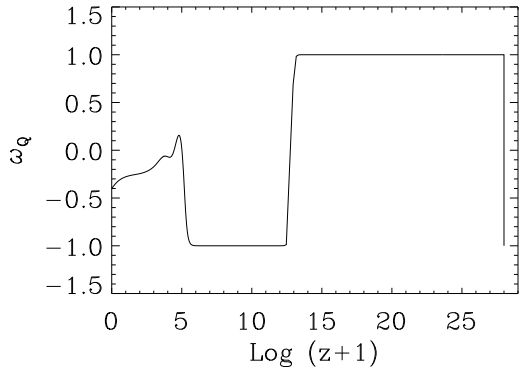}}
\end{center}
\caption{Equation of state in function of the redshift starting 
at equipartition.}
\label{omega0,17}
\end{figure}

The value of $\omega _Q$ today for this model is found to be 
$\omega _Q\approx -0.4$. Therefore it is clear that this case cannot 
be considered as a realistic case but rather as a toy model. 
\par
In order to illustrate the insensitivity to the initial conditions, the 
two following figures show the same case as previously but with 
an initial value of the scalar field given by $Q_{\rm i}\approx 0.2 \times 10^{-4}m_{\rm Pl}$. 

\begin{figure}
\begin{center}
\leavevmode
\hbox{%
\epsfxsize=8cm
\epsffile{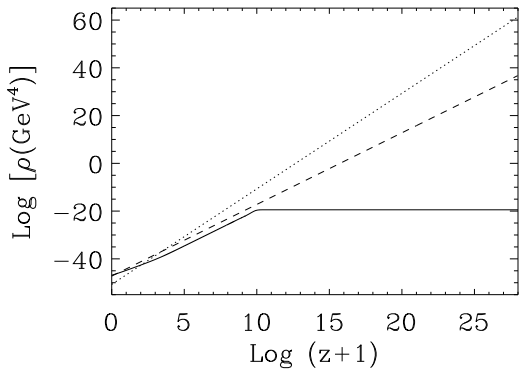}}
\end{center}
\caption{Energy density of radiation, matter and quintessence in function 
of the redshift starting at $Q_{\rm i}\approx 0.2 \times 10^{-4}$.}
\label{rho0,4}
\end{figure}

This case corresponds to an ``undershoot''. One can see that the field starts 
directly from its frozen value. Finally, the evolution of the equation of 
state is displayed. It is apparent that it leads to the same cosmology 
today with $\omega _Q\approx -0.4$

\begin{figure}
\begin{center}
\leavevmode
\hbox{%
\epsfxsize=8cm
\epsffile{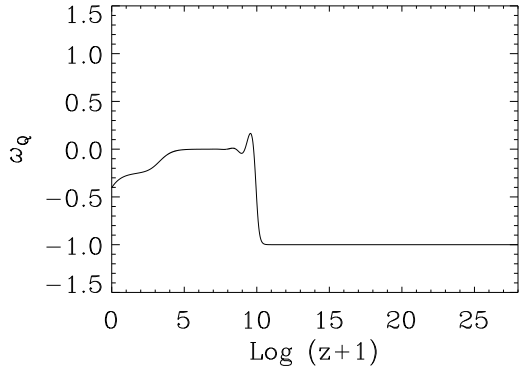}}
\end{center}
\caption{Equation of state for quintessense starting at $Q_{\rm i}\approx 0.2 \times 10^{-4}$.}
\label{omega0,4}
\end{figure}

In the next section we study the influence of the quantum corrections to the 
potential (\ref{invpot}) on the properties described in this section.

\section{Quantum Corrections to non SUSY models}

At the classical level we have chosen a potential given by 
Eq. (\ref{invpot}). However, it is a generic effect that this potential 
will be modified when quantum corrections are taken into account. In this 
section, we only study the one loop corrections. These types of corrections 
automatically cancel out when the model is supersymmetric. Other corrections such as 
the corrections due to curvature effects and to the kinetic terms will  
 be studied in the next sections.
\par
The modified potential reads \cite{CE,W,IIM}:
\begin{eqnarray}
\label{corrpot}
V(Q) &=&\frac{\Lambda ^{4+\alpha }}{Q^{\alpha }} 
+\frac{\Lambda ^4}{32\pi ^2}\ln \frac{\Lambda ^2}{\mu ^2}
+\frac{m^2\Lambda ^2}{32 \pi ^2} \nonumber \\ 
& & +\frac{m^4}{32\pi ^2}\biggl(\ln \frac{m^2}{\Lambda ^2}-\frac{3}{2}\biggr),
\end{eqnarray}
where $\Lambda $ is an effective cut-off already defined and $\mu $ is the natural 
energy scale of the theory. This expansion is obtained by calculating the one 
loop Feynman diagrams. This amounts to evaluating the integral $\int {\rm d}^4p
\ln (p^2+m^2)$ properly regularized by $\Lambda $. This choice has been made 
because $\Lambda $ turns out to be the natural cut-off in the physical models 
considered in this paper \cite{PB}. The energy scale $\mu $ appears in the 
renormalization conditions. It turns out that its precise value is 
not important for our purpose since it only appears in the logarithm. The effective ``mass'' 
$m^2$ is equal by definition to:
\begin{equation}
\label{m2}
m^2 \equiv \frac{{\rm d}^2V(Q)}{{\rm d}Q^2}=
\alpha (\alpha +1)\frac{\Lambda ^{4+\alpha }}{Q^{\alpha +2}}.
\end{equation}
The second term in Eq. (\ref{corrpot}) does not depend on the field $Q$. This 
term will contribute as a cosmological constant. Of course all the other 
fields in the Universe also give contributions to the cosmological constant. It 
is hoped that, by some unknown mechanism, the total contribution vanishes, see 
the introduction. This 
assumption is in the spirit of the quintessential models in which there is no 
need of a cosmological constant in the Einstein equations. For all these reasons we will 
not consider 
the second term in Eq. (\ref{corrpot}) in what follows. 
\par
Introducing the expression giving $m^2$ into the formula of the corrected 
potential, one finds:
\begin{eqnarray}
\label{corrpotQ}
&V&(Q)=\frac{\Lambda ^{4+\alpha }}{Q^{\alpha }} 
+\frac{\alpha (\alpha +1)\Lambda ^{6+\alpha }}{32\pi ^2}\frac{1}{Q^{\alpha +2}}
\nonumber \\
&+& \frac{\alpha ^2(\alpha +1)^2\Lambda ^{8+2\alpha }}{32\pi ^2}
\frac{1}{Q^{2\alpha +4}}
\biggl[\ln \biggl(\frac{\alpha (\alpha +1)\Lambda ^{2+\alpha }}{Q^{\alpha +2}}\biggr)
-\frac{3}{2}\biggr]. 
\end{eqnarray}
We see that the functional form of the potential is no longer the same. 
\par
We now need to estimate the orders of magnitude of the corrections to see whether they can be 
important. As an example, let us consider the case $\alpha =6$ for 
which $\Lambda \approx 4.8 \times 10^{6}$GeV. The first change is that, now, we 
must have $Q_{\rm i} \ge 4 \times 10^{-15}m_{\rm Pl}$ initially in order that 
$\rho _Q \le \rho _B$ at $z=10^{28}$. Interestingly enough this constraint comes from 
the last term in Eq. (\ref{corrpotQ}) which is dominant at this redshift. This means that 
there exists a region for which the quantum corrections are more important 
than the unperturbed potential. As expected, this happens in the early 
Universe, at high energy. A quick estimate enables to show that 
quantum corrections are dominant if $Q_{\rm i} \le 2 \times 10^{-13}m_{\rm Pl}$, i.e. for 
$10^{25}\mbox{GeV}^4 \le \rho _Q \le 10^{61}\mbox{GeV}^4$ initially. Therefore, 
among the $95$ orders of magnitude in which the initial energy density 
of quintessence can vary, $35$ of them are dominated by quantum corrections 
including the most physical case of the equipartition for which 
we have $Q _{\rm i}\approx 6 \times 10^{-15}m_{\rm Pl}$ corresponding to 
$\Omega _{Qi}\approx 10^{-4}$. We conclude that, a priori, quantum corrections 
must be taken into account in any realistic model of quintessence.
\par
However, we find numerically that the final value of $\rho _Q$ and $\omega _Q$ is the 
same with and without 
quantum corrections even if we start from equipartition. As this conclusion is 
not changed if one 
considers other values for $\alpha $, we have demonstrated that quintessence possesses 
another remarkable property: it is stable against quantum corrections. In fact the 
evolution of both the energy density and the equation of state with and without 
the quantum corrections is the same during all the cosmic evolution. This is due to 
the fact that at the beginning of the evolution the field rolls down the potential 
very quickly and leaves the region where quantum corrections are important in a very 
short time.
\par
In conclusion  we have shown that non SUSY models of quintessence 
are stable against one loop quantum corrections to the effective potential. This property 
is generic 
and does not depend on the precise value of $\alpha $.
\par
In the next section, we start examining SUSY models of quintessence. In that 
case the quantum corrections studied in this section automatically cancel 
out. 

\section{SUSY models}

As seen in the previous sections the quintessence field varies over a large range
of values, as high as the Planck scale. It is therefore compulsory to treat the 
quintessence behaviour within the framework of models encapsulating the expected behaviour 
of high energy physics. We will use supersymmetric models.
One of the advantages of these models is their stability with respect to quantum 
corrections. In particular it is known that the non-renormalisation theorem preserves 
superpotentials, there is only a wave function renormalisation.

\subsection{SUSY gauge theories}

The potential in $1/Q^{\alpha}$ which leads to quintessence has a
natural supersymmetric origin. This was first shown in Ref. \cite{PB}. The aim 
of this section is to generalise the models investigated in Ref. \cite{PB}. This type 
of potential is generated at low energy
$E\le \Lambda$ in supersymmetric gauge theories due to non-perturbative
effects along flat directions of the scalar potential.
In this section we present the general features of supersymmetric gauge theories.
More details can be found in \cite {skiba,PBsusy}
\par
Before dealing with the general case let us recall the most famous example
constructed with the gauge group $SU(N_{\rm c})$ and $N_{\rm f}$ quarks and antiquarks  
$(z^i_{\alpha },\bar z^i_{\alpha }),\ i=1\dots N_{\rm f}$ and 
$\alpha=1\dots N_{\rm c},$ in the fundamental and 
antifundamental representations of the gauge group. The dynamics of this model is governed 
by the renormalisation evolution of the gauge coupling 
constant. For $b=3N_{\rm c}-N_{\rm f}>0$
the gauge coupling constant becomes strong at low energy. In this infrared regime
it is relevant to study the configurations of the scalar components of the quarks and 
antiquarks which have zero energy. The potential $V_{\rm D}=g^2 D^2/2$ vanishes when the 
$D$-terms $D^a=z^{i*}_{\alpha}T^a_{\alpha\beta}z^i_{\beta} 
-\bar z^i_{\alpha} T^a_{\alpha\beta}\bar z^i*_{\beta}$ vanish leading to:
\begin{equation}
z^i_{\alpha }z^{i*}_{\beta }-\bar z^{i*}_{\alpha }\bar z^i_{\beta }
=\lambda {\rm \delta }_{\alpha \beta}.
\end{equation}
The manifold of solutions of these equations is called the moduli space of the 
gauge theory. The moduli space is in one to one correspondence with the gauge invariant 
polynomials $I^a$ via the equation
\begin{equation}
\frac{{\rm \partial }I^a}{{\rm \partial }z^i_{\alpha }}=z^{i*}_{\alpha }.
\end{equation}
In the case $N_{\rm f}<N_{\rm c}$ there is just one gauge 
invariant  $I^a\equiv M^{ij}= z^{i}_{\alpha }\bar z^j_{\alpha }$ called the meson 
field. The low energy dynamics is expressed in terms of the meson 
field. As the gauge group becomes strongly interacting at low energy  
$E\le \Lambda $ where $\Lambda $ is the strong interaction scale, non-perturbative effects 
due to the condensation of gauginos lead to a non-zero superpotential for the meson 
field,
\begin{equation}
W\propto \frac{\Lambda^{b/(N_{\rm c}-N_{\rm f})}}{{(\det M)}^{1/(N_{\rm c}-N_{\rm f})}}.
\end{equation}
This potential is deduced for $N_f=N_c-1$ by an instanton calculation and for 
all $N_{\rm f}$ using the decoupling technique.
Let us now focus on the amplitude mode $M^{ij}=Q^2\delta^{ij}$ obtained
from $z^i=Q z^i_0,\bar z^i=Q\bar z^i_0$. Note that the case of two field directions 
has recently been investigated in Ref. \cite{MPR}. Starting from a flat K\"ahler potential
$K=\hbox {tr} (zz^{\dagger} +z\bar z^{\dagger})$ the classical K\"ahler potential at low 
energy becomes $K=QQ^*$ after normalizing 
$\hbox {tr} (z_0z^{\dagger}_0+\bar z_0\bar z^{\dagger}_0)=1$.
This yields the low energy scalar potential for the amplitude mode
\begin{equation}
V(Q)=\frac{\Lambda^{2b/(N_{\rm c}-N_{\rm f})}}{Q^{2(N_{\rm c}+N_{\rm f})/(N_{\rm c}-N_{\rm f})}}.
\end{equation}
From the previous equation, it is clear that the so far arbitrary coefficient $\alpha $ is 
now given by:
\begin{equation}
\alpha=2\frac{N_{\rm c}+N_{\rm f}}{N_{\rm c}-N_{\rm f}},
\end{equation}
which is always greater than two. This result had already been obtained 
in Ref. \cite{PB}. The analysis carried out in this particular model is in fact typical 
of most supersymmetric gauge theories. We now show how this approach can be generalized.
\par
Consider a gauge group $G$ and the matter fields $z^i$ in the representation
$R_{\rm i}$ of the gauge group. Define the Dynkin index $\mu_{\rm i}$ of a representation 
$R_{\rm i}$ to be
\begin{equation}
\hbox {tr}(T^aT^b)=\frac{\mu_{\rm i}}{\theta^2}\delta^{ab},
\end{equation}
where the $T^a$'s are the Hermitean generators of the  
representation $R_i$ and $\theta$ is the long root of the Lie algebra 
of $G$. For instance for $G=SU(N_{\rm c})$, the previous results are retrieved using
$\mu_{\rm f}=1$ for the fundamental representation and $\mu_{\rm adj}=2N_{\rm c}$ for the 
adjoint representation. The beta function determining the evolution of the gauge coupling 
constant depends on $2b=3\mu_{\rm adj}-\mu$ where $\mu=\sum_i \mu_i$ is the sum of all the 
Dynkin indices of the matter fields. The gauge theory is asymptotically free and strongly 
coupled in the infrared when $b>0$. At low energy along the flat directions determined by 
the vanishing of the $D$ terms the dynamics of the gauge theory is encoded in the properties 
of the polynomial gauge invariants $I^a$. When the Dynkin indices 
satisfy $\mu<\mu_{\rm adj}$ the ring of gauge invariants is free. In terms of these gauge 
invariants the strong non-perturbative dynamics
of the gauge theory generates a superpotential $W(I)$ whose form is dictated by symmetry 
arguments. This superpotential when expressed in terms of the original
matter fields $z^i$ reads
\begin{equation}
W(Q)\propto\frac {\Lambda^{b/(\mu_{\rm adj}-\mu )}}{\biggl(\prod_i (z^i)^{\mu_i}\biggr)^
{2/(\mu_{\rm adj}-\mu )}}.
\end{equation}
As for the $SU(N_{\rm c})$ case one is interested in the amplitude mode $Q$ whose 
classical K\"ahler potential is flat. This leads to the scalar potential:
\begin{equation}
V(Q)=\frac{\Lambda^{2b/(\mu_{\rm adj}-\mu )}}{Q^{2(\mu _{\rm adj}+\mu )/(\mu_{\rm adj}-\mu )}},
\end{equation}
where the exponent $\alpha$ is now given by the following expression:
\begin{equation}
\alpha=2\frac{\mu _{\rm adj}+\mu}{\mu_{\rm adj}-\mu}.
\end{equation}
which is always greater than two.
\par
Therefore, the inverse power potential is a generic prediction of supersymmetric gauge 
theories. Nevertheless it is dependent on the hypothesis that the K\"ahler potential is 
flat. This is plausible when $Q\ge \Lambda$ but needs to be reconsidered below this 
scale. This leads to the dangerous K\"ahler corrections which will be studied later. It has also 
been assumed that spacetime was flat. Curvature corrections are studied in the next section.
 
\subsection{Curvature corrections}

In the previous section we have treated the globally supersymmetric case as if the spacetime 
was not curved. These curvature effects are usually neglected as the typical scale $H$ is 
too small compared to the particle physics scales (recall 
that $H_0 \approx 2.1h\times 10^{-42}\mbox{GeV}$). Concerning quintessence this assumption 
has to be carefully checked as Eq. (\ref{attractor}) implies that the effective mass of 
the quintessence field is also of 
order $H$. This requires to study in a painstaking fashion the effects of curvature on 
global supersymmetry. A general argument is presented in an appendix while an explicit 
calculation is given here.
\par
We assume that the quintessence field belongs to a chiral supermultiplet, i.e. a 
complex scalar field and a Weyl spinor. We shall first examine the case of a free field 
showing that global supersymmetry is broken explicitly by curvature effects in a FLRW 
spacetime. This is due to the fact that the bosonic particles are created from the gravitational 
background whereas this is not the case for the fermions since they are conformally 
invariant. In the interacting case the breaking of supersymmetry by the curvature follows 
from a supergravity argument. In that case we evaluate the effective action at the one-loop 
level using the zeta function technique. 

\subsubsection{Complex scalar field}

The action of the complex scalar field $\varphi (x^{\mu})$ is given by the 
following expression:
\begin{equation}
\label{defSscalar}
S_{\varphi }=-\int {\rm d}^4x\sqrt{-g}g^{\mu \nu}
{\rm \partial }_{\mu }\varphi {\rm \partial }_{\nu }\varphi ^*.
\end{equation}
It is convenient to separate the real and imaginary parts of the field 
and to write:
\begin{equation}
\label{ReIm}
\varphi =\frac{1}{\sqrt{2}}(\varphi _1+i\varphi _2), \quad 
\end{equation}
where $\varphi _i$, $i=1,2$, are now real scalar fields. Since the spatial sections are flat, the 
fields can be Fourier decomposed. It is convenient to perform this decomposition according to:
\begin{equation}
\varphi _i(\eta ,{\bf x})=\frac{1}{a(\eta )}\frac{1}{(2\pi )^{3/2}}
\int {\rm d}^3{\bf k}\mu _i(\eta ,{\bf k})e^{i{\bf k}\cdot {\bf x}},
\end{equation}
where we have extracted a factor $1/a(\eta )$ in the time dependent amplitude 
of the Fourier component for future convenience. The Fourier components 
$\mu _i(\eta ,{\bf k})$ are such that $\mu _i(\eta ,{\bf k})
^*=\mu _i(\eta ,-{\bf k})$ because the fields $\varphi _i$ are real. It can be easily 
seen that $\mu _i(\eta ,{\bf k})$ obeys the equation of a parametric oscillator:
\begin{equation}
\label{eqofmscalar}
\mu _i(\eta ,{\bf k})''+\biggl(k^2-\frac{a''}{a}\biggr)\mu _i (\eta ,{\bf k})=0.
\end{equation}
This equation reduces to the equation of an ordinary harmonic oscillator in a 
Minkowski or radiation dominated Universe where $a''=0$.
\par
When quantization is carried out the fields $\varphi _i$ become operators. In the 
canonical approach, the complex scalar field can be expressed as:
\begin{eqnarray}
\label{complexfieldoperator}
\hat{\varphi }(\eta ,{\bf x}) &=& 
\frac{1}{a(\eta )}
\frac{1}{(2\pi )^{3/2}} \nonumber \\
&\times & \int \frac{{\rm d}^3{\bf k}}{\sqrt{2k}}
[d_{1{\bf k}}(\eta )e^{i{\bf k}\cdot {\bf x}}
+d_{2{\bf k}}^{\dag}(\eta )e^{-i{\bf k}\cdot {\bf x}}],
\end{eqnarray}
where the operators $d_{i{\bf k}}$ are related to the operators of creation and 
annihilation $c_{i{\bf k}}(\eta )$ and $c_{i{\bf k}}^{\dagger }(\eta )$ associated 
to the field operators $\hat{\varphi }_i$ and satisfying the commutation 
relation $[c_{i{\bf k}}(\eta ),c_{j{\bf k}'}^{\dag}(\eta )]= \delta _{ij}\delta ({\bf k}
-{\bf k}')$ through the expressions:
\begin{equation}
\label{defd1d2}
d_{1{\bf k}}\equiv \frac{c_{1{\bf k}}+ic_{2{\bf k}}}{\sqrt{2}}, \quad 
d_{2{\bf k}}^{\dag} \equiv \frac{c_{1{\bf k}}^{\dag }+ic_{2{\bf k}}^{\dag}}{\sqrt{2}}.
\end{equation}
The expression (\ref{complexfieldoperator}) can be used in order to express the 
Hamiltonian operator for the complex 
scalar field in terms of creation and annihilation operators. The result reads:
\begin{eqnarray}
\label{Hoperator}
H &=& \frac{1}{2}\sum _{i=1}^2\int {\rm d}^3{\bf k}\biggl(k
(c_{i{\bf k}}c_{i{\bf k}}^{\dag }+
c_{i,-{\bf k}}^{\dag }c_{i,-{\bf k}}) \nonumber \\
&-&i\frac{a'}{a}(c_{i{\bf k}}c_{i,-{\bf k}}
-c_{i{\bf k}}^{\dag }c_{i,-{\bf k}}^{\dag })\biggr).
\end{eqnarray}
The first term of the Hamiltonian represents the Hamiltonian of a collection of 
harmonic oscillators. The second term represents the interaction between the classical 
background and the quantum field. It is proportional to the first derivative 
of the scale factor and therefore vanishes in the Minkowski case. This term is 
responsible for the creation of (pair of) particles. If we start from the 
vacuum state $|0\rangle $ (no particle), then due to the presence of this interaction 
term, the state will evolve in a vacuum squeezed state \cite{G}. 

\subsubsection{Weyl spinor field}

The action of the Weyl spinor field is given by the following expression:
\begin{eqnarray}
\label{actionspinor1}
S_{\psi } &=& -\frac{1}{2}\int {\rm d}^4x\sqrt{-g}i\biggl(\bar{\psi }_{\dot{\alpha }}
(\bar{\Sigma }^{\mu })^{\dot{\alpha }\alpha }{\rm D}_{\mu }\psi _{\alpha } \nonumber \\
&-& ({\rm D}_{\mu }\bar{\psi }_{\dot{\alpha }})
(\bar{\Sigma }^{\mu })^{\dot{\alpha }\alpha }\psi _{\alpha }\biggr).
\end{eqnarray}
In this expression $\Sigma ^{\mu }$ are the Pauli matrices in a FLRW Universe and 
${\rm D}_{\mu }$ denotes the covariant derivative for a Weyl spinor. In the 
following, we specify more the definitions used here. The Dirac matrices in curved 
spacetime, $\Gamma ^{\mu }$, are defined according to the equation
$\{\Gamma ^{\mu },\Gamma ^{\nu}\}=-2g^{\mu \nu}$. As a consequence, these matrices 
can be expressed 
in terms of the vierbein $e^a{}_{\mu }$ defined 
by $g_{\mu \nu }\equiv e^a{}_{\mu }e^b_{\nu }\eta _{ab}$, where 
$\eta _{ab}$ is the Minkowski metric. This results in the 
equation $\Gamma ^{\mu }=e^{\mu }{}_a\gamma ^a$ where the $\gamma ^a$ 
are the Dirac matrices in flat Minkowski spacetime. In the case of a FLRW Universe, 
one has $g_{\mu \nu}=a^2(\eta )\eta _{\mu \nu}$ which means that the veirbein are 
given by $e^a{}_{\mu }=a(\eta )\delta ^a{}_{\mu }, 
e^{\mu }{}_a=[1/a(\eta )]\delta ^{\mu }{}_a $. This implies that the FLRW Dirac matrices 
can be written as $\Gamma ^{\mu }=[1/a(\eta )]\gamma ^{\mu }$. Since in the Weyl 
representation the Dirac matrices are anti-diagonal and can be expressed as:
\begin{equation}
\label{WrepGamma}
\Gamma ^{\mu }=\left(
\begin{array}{cc}
0 & \Sigma ^{\mu } \\
\bar{\Sigma }^{\mu } & 0
\end{array}
\right),
\end{equation}
where $\Sigma ^{\mu }$ are the Pauli matrices in a FLRW spacetime, one reaches the 
conclusion that $\Sigma ^{\mu }=[1/a(\eta )]\sigma ^{\mu }$ where $\sigma ^{\mu }$ are 
the Pauli matrices in Minkowski spacetime. Let us now turn to the expression of the 
covariant derivative. For a four-dimensional Dirac spinor $\Psi _{\rm D}$ which can 
be written in the Weyl representation as
\begin{equation}
\label{Diracspin}
\Psi _{\rm D}=\left(
\begin{array}{cc}
\psi _{\alpha } \\
\bar{\chi }^{\dot{\alpha }}
\end{array}
\right),
\end{equation}
the covariant derivative can be expressed through the two expressions:
\begin{eqnarray}
\label{nablaWeyl1}
{\rm D}_{\mu }\psi _{\alpha } &=& {\rm \partial }_{\mu }\psi _{\alpha }
+\frac{1}{2}w_{ab\mu }(\sigma ^{ab})_{\alpha }{}^{\beta }\psi _{\beta }, \\
\label{nablaWeyl2}
{\rm D}_{\mu }\bar{\psi} _{\dot{\alpha }} &=&
{\rm \partial }_{\mu }\bar{\psi }_{\dot{\alpha }}
- \frac{1}{2}w_{ab\mu }
(\bar{\sigma }^{ab})^{\dot{\beta }}{}_{\dot{\alpha }}\bar{\psi }_{\dot{\beta }}.
\end{eqnarray}
The matrices $\sigma ^{ab}$ and $\bar{\sigma }^{ab}$ are defined according to 
the usual expressions, namely $\sigma ^{ab}\equiv (1/4)(\sigma ^a\bar{\sigma }^b
-\sigma ^b\bar{\sigma }^a)$ and $\bar{\sigma }^{ab}\equiv (1/4)(\bar{\sigma }^a\sigma ^b
-\bar{\sigma }^b\sigma ^a)$ \cite{WB}. In a FRLW Universe, the components of the spin connection 
can be written as:
\begin{eqnarray}
\label{w0}
w_{000} &=& w_{0i0}=w_{i00}=w_{ij0}=0, \\
\label{wi}
w_{00i} &=& w_{mni}=0, \quad w_{j0i}=-w_{0ji}=\frac{a'}{a}\delta _{ij}.
\end{eqnarray}
Let us now redefine the spinor $\psi _{\alpha }$ by $\psi _{\alpha }
\equiv \chi _{\alpha }/a(\eta )^{3/2}$. Then 
the Lagrangian ${\cal L}$ given in Eq. (\ref{actionspinor1}) can be re-written as:
\begin{equation}
\label{Lchi}
{\cal L}=-\frac{i}{2}\biggl( \bar{\chi }_{\dot{\alpha }}(\bar{\sigma }^{\mu })
^{\dot{\alpha }\alpha }{\rm \partial }_{\mu }\chi _{\alpha }
-({\rm \partial }_{\mu }\bar{\chi }_{\dot{\alpha  }})(\bar{\sigma }^{\mu })^
{\dot{\alpha }\alpha }\chi _{\alpha }\biggr).
\end{equation}
This equation shows that the a spinor field in FLRW spacetime is conformally 
invariant \cite{Parker}. Contrary to the case of a scalar field, there is no interaction term 
between the background geometry and the quantum field. As a consequence there is 
no creation of (massless) fermionic particles. Another way to put it is to say 
that if we start from the vacuum state, the system will remain in this state 
for ever. At this level, we could already conclude that SUSY is broken 
in a FLRW spacetime. Indeed the previous phenomenon is equivalent to say that 
the bosons possess a time dependent mass whereas the fermions still have a constant 
mass. Therefore, the condition $m_{\rm B}=m_{\rm F}$ can no longer be satisfied and 
global SUSY must be broken. The phenomenon of creation of particles in curved spacetimes 
is responsible for the SUSY breaking.
\par
In order to demonstrate explicitly how this property shows up, let us 
pursue the calculations in more details. Having seen that the fermions 
$\chi _{\alpha }$ behaves like free
fermions we can canonically quantize the fermionic field. To do
so we need to define the zero modes of the Dirac operator acting
on Weyl fermions
\begin{eqnarray}
u_{\alpha }^+({\bf k}) &=& \frac{-i}{\sqrt{k_0+k_3}}  
\left(\begin{array}{cc}
-k^{-}  & \\
k_0+k_3 & \\
\end{array}
\right), \\
u_{\alpha }^-({\bf k}) &=& \frac {i}{\sqrt{k_0-k_3}}
\left(
\begin{array}{cc}
k^{-}  & \\
k_0-k_3 & \\
\end {array}
\right),
\end{eqnarray}
where $k^{\pm }\equiv k_1\pm ik_2$. Then, the fermionic field operator 
$\hat{\chi }_{\alpha }$ can be canonically decomposed according to:
\begin{eqnarray}
\label{fermionope}
\hat{\chi }_{\alpha }(\eta ,{\bf x}) &=& \frac{1}{(2\pi)^{3/2}}\int 
\frac{{\rm d}^3{\bf k}}{\sqrt {2k_0}}e^{i{\bf k}\cdot {\bf x}} \nonumber \\
& &\times \biggl(b_{+,{\bf k}}u_{\alpha }^+({\bf k})e^{ik_0\eta } +b_{-,{\bf k}}
u_{\alpha }^-({\bf k})e^{-ik_0\eta}\biggr),
\end{eqnarray}
where the creation and annihilation operator satisfy the anticommutation 
relation
\begin{equation}
\{b_{+,{\bf k}}, b_{+,{\bf k}'}^{\dagger}\}=\delta ({\bf k}-{\bf k}'), \quad 
\{b_{-,{\bf k}}, b_{-,{\bf k}'}^{\dagger}\}=\delta ({\bf k}-{\bf k}').
\end{equation}
We are now in position to study the explicit breaking of
supersymmetry due to the curvature of the FLRW spaces. This is the aim of the 
next section.

\subsubsection{SUSY breaking in curved spacetime}

The free Lagrangian in the rescaled bosonic
and fermionic is a free supersymmetric Lagrangian apart explicit
breaking terms proportional to $a'/a$. The supersymmetric current is 
no longer conserved leading to a time dependent supersymmetric 
charge. A natural definition of the current is provided by the following 
expression which involves only the rescaled fields:
\begin{equation}
J_{\alpha}=\chi_{\alpha}\partial_0 (a\varphi ) -a\varphi \partial_0 \chi_{\alpha}.
\end{equation}
Accordingly, the supersymmetric charge can be expressed as:
\begin{equation}
Q_{\alpha}=\int d^3 x J_{\alpha}.
\end{equation}
The supersymmetric charge can be expressed in terms of the
canonical creation and annihilation operators. Using Eqns. 
(\ref{complexfieldoperator}) and (\ref{fermionope}), this leads to the relation
\begin{eqnarray}
& &\frac{{\rm d}Q_{\alpha}}{{\rm d}\eta}=\frac{a''}{a}\int \frac{{\rm d}^3{\rm k}}{2k_0}
\nonumber \\
& & \biggl(u_{\alpha }^+({\bf k})b_{+,{\bf k}}d_{1{\bf k}}^{\dagger }
+u_{\alpha }^+(-{\bf k})b_{+,-{\rm k}}d_{2{\rm k}}e^{ik_0\eta} \nonumber \\
& & +u_{\alpha }^-({\bf k})b_{-,{\bf k}}d_{1{\bf k}}^{\dagger}
+u_{\alpha }^-(-{\bf k})b_{-,-{\rm k}}d_{2{\rm k}}e^{-ik_0\eta}\biggr). 
\end{eqnarray}
Recalling that the scalar of curvature is given by $R=6a''/a^3$ we have obtained that
supersymmetry is explicitly broken by curvature effects. Notice
that when $R=0$ supersymmetry is preserved. This corresponds to
either Minkowski spacetime or the radiation dominated FLRW spacetime for 
which one has $a(\eta )\propto \eta $. This results is not particular to the 
non-interacting theory but can be generalised  by considering $N=1$ supergravity in 
four dimensions. In the low energy limit when non-renormalisable gravitational 
interactions are neglected, i.e. $m_{\rm Pl}\to \infty$, the supergravity Lagrangian reduces 
to the curved action that we have considered previously. Supersymmetry is preserved when 
the background metric of the curved spacetime possesses enough Killing 
spinors, i.e. solutions of the spinorial equation
\begin{equation}
{\rm D}_{\mu }\epsilon=0,
\end{equation}
where ${\rm D}_{\mu }$ is the covariant derivative and $\epsilon $ is the supersymmetry 
variation parameter. By considering $[{\rm D}_{\mu }, {\rm D}_{\nu }]\epsilon=0$ and the 
maximal symmetry of the FLRW metric one finds that supersymmetry is only preserved for flat 
FLRW spacetimes, i.e. $R=0$. Since this is true even in the presence of interacting chiral 
superfields, this generalises the previous result.
\par
The fact that SUSY is broken in a curved spacetime will induce corrections to the 
quintessential potential since this one is no longer protected. We now evaluate the 
order of magnitude of these 
corrections. We have derived in the appendix the one-loop effective potential in the 
interacting case in the presence of curvature effects. The one loop effective 
potential reads:
\begin{eqnarray}
\label{corrcur}
{\rm \delta }V_{\rm eff} (Q,\eta ) &=& \frac {1}{32\pi^2 }
\biggl\{ \biggl[ \vert m\vert ^2 -\frac{R}{6}\biggr]^2
\biggl[\ln \biggl(\frac{\vert m\vert ^2-R/6}{\mu^ 2}\biggr)-\frac {3}{2}\biggr]
\nonumber \\
&-& \biggl[\vert m\vert ^2 +\frac{R}{12}\biggr]^2
\biggl[\ln \biggl(\frac{\vert m\vert ^2+R/12}{\mu ^2}\biggr)-\frac {3}{2}\biggr]\biggr\},
\end{eqnarray}
where $\mu $ is the renormalization scale. As for the quantum corrections, the 
dependence in $Q$ appears through the 
relation $m\equiv m_{\rm B}=m_{\rm F}={\rm d}^2V(Q)/{\rm d}Q^2$. In addition 
there is now an explicit time dependence due to the presence of the scalar of 
curvature in Eq. (\ref{corrcur}). The evolution starts during the radiation 
dominated epoch where global SUSY is preserved since $R=0$. As a consequence, 
Eq. (\ref{corrcur}) implies that ${\rm \delta }V_{\rm eff} (Q,\eta )=0$, i.e. 
no corrections are generated. Notice 
that in contrast with the quantum corrections the curvature effects manifest 
themselves at later times. In the matter dominated era, the effects of curvature no 
longer vanishes and the corrections modify the power law potential. These corrections 
are the usual quantum corrections plus additional contributions proportional 
to $H^2$. It has been shown in the previous section that quantum corrections are 
important only deep in the radiation dominated era. Since $H^2$ is a tiny number 
during the matter dominated era, we conclude that the curvature effects for 
quintessence are therefore negligible.

\subsection{K\"ahlerian corrections}

In gauged supersymmetric models where flat directions are lifted non-perturbatively
below a scale $\Lambda$ where the dynamics of the gauge group becomes strongly coupled, we 
have seen that the effective superpotential for the amplitude mode
$Q$ has the required form to lead to a inverse power law quintessential scalar 
potential provided the K\"ahler potential is flat. The non-renormalisation theorem
guarantees that the superpotential does not receive quantum corrections due to
radiative corrections. This is not the case of the K\"ahler potential which is not 
protected, and therefore is modified at low energy. The low energy K\"ahler potential 
becomes a complicated function $K(Q,Q^*;\Lambda)$ which can be
expanded in Taylor series as:
\begin{eqnarray}
\label{TaylorK}
K(Q,Q^*;\Lambda) &=& QQ^* \nonumber \\
&+&\biggl(\sum_{\stackrel{m>0,n>0}{m+n>2}} 
a_{mn}\frac{Q^mQ^{*n}}{\Lambda^{m+n-2}}+ {\rm c.c.}\biggr).
\end{eqnarray}
This expansion is valid as long as $Q\le \Lambda $ which means that the K\"ahler 
potential will be modified at the beginning of the evolution, deep in the 
radiation dominated era. As already mentioned, when $Q\ge \Lambda $, the K\"ahler 
potential can be considered as flat. In this respect the situation is similar to 
what happens for the quantum corrections studied previously. 
In the case of the $SU(N_{\rm c})$ gauge theory with $N_{\rm f}$ flavour there is a 
useful ansatz for the K\"ahler potential which illustrates the K\"ahler 
correction. Indeed a K\"ahler potential of the form 
$K=(1/N_{\rm f})\mbox{tr} (MM^{\dagger}+\Lambda^4)^{1/2}$ leads to an 
expansion $K\approx (QQ^{\dagger})^2/(2\Lambda^2) - (QQ^{\dagger})^4/(8\Lambda^6)
+\cdots$ when $Q\ll \Lambda$ and $K\approx QQ^{\dagger}$ when $Q\gg \Lambda$.
This is typical of the situation we are going to describe.
\par
Let recall the structure of the low energy Lagrangian 
in the presence of a non-flat K\"ahler potential. In a flat 
gravitational background the Lagrangian reads 
\begin{equation}
{\cal L}=\int {\rm d}^4 x\biggl(g{\rm \partial} Q{\rm \partial } Q^* -\frac{1}{g}\biggl\vert 
\frac{{\rm \partial} W}{{\rm \partial} Q}\biggr\vert^2\biggr),
\end{equation}
where we have denoted by $Q$ the chiral superfield whose scalar component 
is $Q$. The metric on the one dimensional complex curve defined by $K$ is given by 
the following expression
\begin{equation}
g={\rm \partial }_Q{\rm \partial }_{Q^*}K.
\end{equation}
Using the Taylor expansion of the K\"ahler potential given 
by Eq. (\ref{TaylorK}) and the previous relation, one can easily deduce 
the expression of the K\"ahler metric:
\begin{equation}
g=1+\sum_{\stackrel{m>0,n>0}{m+n>2}}mna_{mn} \frac{Q^{m-1}Q^{*(n-1)}}{\Lambda^{m+n-2}}.
\end{equation}
Notice that the flat metric is modified by the quantum corrections in $1/\Lambda $. The 
scalar Lagrangian can be rearranged in order to render the physical meaning
of the non-trivial K\"ahler potential more transparent. Let us redefine the fields according to
\begin{equation}
\frac{{\rm d}\tilde Q}{{\rm d}Q}=\sqrt{g},
\end{equation}
where we are now specializing our results to the real case $Q=Q^*$.
The square root of $g$ can be expanded in a 
series $\sqrt{g}=1+\sum_{n>1} b_n(Q^n/\Lambda ^n)$
leading, after an inversion which can be carried out inductively, to:
\begin{equation}
Q=\tilde Q+\sum_{n>1} c_n \frac{\tilde Q^{n+1}}{\Lambda^n}.
\end{equation}
This redefinition transforms the kinetic terms of $Q$ into canonically normalized ones 
for $\tilde{Q}$, namely ${\rm \partial }^{\mu }\tilde{Q}{\rm \partial }_{\mu }
\tilde{Q}$. On the other hand, the scalar potential becomes then
\begin{equation}
V(Q)=\frac {1}{g[Q(\tilde Q)]}\frac{\Lambda^{4+\alpha}}{Q^{\alpha}(\tilde Q)},
\end{equation}
where we have explicitly shown the dependence of $Q$ on $\tilde Q$.
The metric $g$ depends on the new field $\tilde Q$ and is expandable
in a series $g=1+\sum_{n\ge 1} d_n (\tilde Q^n/\Lambda^n)$. As a consequence the scalar 
potential becomes now an infinite series the expression of which is given by
\begin{equation}
\label{potK}
V(Q)=\frac{\Lambda^{4+\alpha}}{\tilde Q^{\alpha}}\biggl(1+\sum_{n>1} e_n 
\frac{\tilde Q^n}{\Lambda^n}\biggr),
\end{equation}
where the coefficients $e_n$ are easily computed inductively. This potential must 
be seen as the effective potential of quintessence in the region 
where $Q\le \Lambda $. This potential has the usual quintessential inverse power behaviour 
but is corrected by positive powers of $\tilde Q$. Whether the properties of the tracking 
solutions will be preserved crucially depends on the coefficients $e_n$. 
\par
Let us conclude this section by stressing that building a realistic model of quintessence 
based on global SUSY appears to be a difficult task. Even if these 
models are free from quantum corrections to the potential and from curvature corrections, the 
tracking properties could be destroyed by the K\"ahlerian corrections as 
shown by Eq. (\ref{potK}). Nevertheless the main difficulty 
is that at the end of the evolution $Q\approx m_{\rm Pl}$, rendering the SUGRA corrections 
unavoidable. We cannot neglect contributions in the scalar potential which are suppressed 
by the Planck mass. Therefore, any realistic model of quintessence must be based on 
SUGRA. The aim of the next section is to study such models.

\section{supergravity models}

\subsection{SUGRA corrections}

In this section we consider the SUGRA version of the model considered 
previously with a superpotential of the 
form $W\propto 1/Q^{\alpha }$. At tree level the supergravity scalar potential 
depends on the potential $G\equiv \kappa K+ \ln (\kappa^3 \vert W\vert^2)$ where $K$ is the
K\"ahler potential and $W$ the superpotential. The scalar potential
is given by:
\begin{equation}
V=\frac{e^{G}}{\kappa ^2}(G^iG_i-3)+V_{\rm D},
\end{equation}
where the indices have been raised using the metric $G_{i\bar
\j}\equiv \partial_i\partial_{\bar \j} G$ and where the derivatives have been
taken with respect to the scalar fields. The term $V_{\rm D}$ comes from the 
gauge sector of the theory and is always positive. In the quintessence
context there is only one field and the K\"ahler potential is
chosen to be flat $K=QQ^*$. This leads to the following scalar potential:
\begin{equation}
V=e^{\kappa Q^2}\frac{\Lambda^{4+\alpha}}{Q^{\alpha}}\biggl(\frac{(\alpha-2)^2}{4}-(\alpha+1)
\kappa Q^2 +\kappa^2Q^4\biggr).
\end{equation}
There are a few remarks at hands about this potential. The first term corresponds to the 
global supersymmetry scalar potential while the other two terms are supergravity 
corrections in $Q/m_{\rm Pl}$. There are two important effects of the supergravity 
corrections. The exponential term in the scalar potential introduces positive
powers of $Q$ of arbitrary degrees. Fortunately this only becomes
relevant for $Q$ of order of the Planck mass, i.e. for the
red-shift $z\approx 0$. More importantly the potential can become
negative due to the second term. This implies that the model can
become non-sensical at the end of the evolution when $\kappa
Q\approx m_{\rm Pl}$. In the case $\alpha=6$, we have checked numerically that this is indeed 
the case. For higher values of $\alpha$ this is still the case proving that globally 
supersymmetric models with an inverse power law superpotential do not resist the supergravity 
corrections.   
\par
The appearance of dangerous negative contributions to the
potential is not accidental, it stems from the $-3$ term in the potential.
A possible way out is to impose that the scalar potential exactly
vanishes and that the scalar potential is entirely due to a
non-flat K\"ahler potential. This is what we are going to study in
the next section.

\subsection{Moduli Quintessence}

Let us consider a supergravity model where there are two types of
fields, the quintessence field $Q$ and matter fields $(X,Y^i)$.
We assume that the gauge group of the model is broken along flat
directions of the $D$ terms such that $X\ne 0, Y^i=0$ where $V_{\rm D}=0$. As 
explained
in the previous section we impose that the scalar potential is
positive to prevent any negative contribution to the energy 
density. This is achieved by considering that
\begin{equation}
\langle W \rangle=0, 
\end{equation}
when evaluated along the flat direction. Moreover we assume that
one of the gradients of the superpotential $W_Y$ does not vanish.
With these assumptions the scalar potential becomes
\begin{equation}
V=e^{\kappa K}K^{YY^*}\vert W_Y\vert^2.
\label{pot}
\end{equation}
As expected the scalar potential is positive and becomes a
function of the quintessence field $Q$ only. The quintessence
property is achieved if this potential possesses the run-away behaviour
of the quintessence field to infinity. 
\par
In the following we shall use 
string-inspired models with an
anomalous $U(1)_X$ gauge symmetry \cite{DSW,IL}. In the context of the heterotic string theory 
it is natural to identify the quintessence field with one of the moduli of the 
string compactification. Indeed the values of the moduli naturally goes to infinity.
This is usually called the moduli problem which in fact turn into a blessing in the 
context of quintessence. Consider the compactification of the weakly coupled 
heterotic string on a Calabi-Yau manifold. The compactification depends explicitly
on the moduli $T_{\alpha}, \alpha=1\dots 3$, representing
deformations of the K\"ahler class of the Calabi-Yau manifold. We
shall concentrate on one moduli $T_{\alpha}=Q$ on which the
modular group $SL(2,Z)$ acts as $Q\to (aQ-ib)/(icQ+d)$ with $ad-bc=1$.
Moreover we suppose that the gauge group factorises as $G\times
U(1)_X$ where $G$ contains the standard model gauge group and
$U(1)_X$ is an anomalous Abelian symmetry. The fields of the
model split into three groups, the field $X$ has a charge $1$
under $U(1)_X$ and is neutral under $G$, the field $Y$ is a
matter field neutral under $G$ and of charge $-2$ under $U(1)_X$ 
while the matter fields $Y_i$ are charged under $G$ and possess 
charges $q_i\ne -2$ under $U(1)_X$.  Under the modular group the matter
fields transform with a modular weight $-1$, i.e. $Y_i\to
(icQ+d)^{-1}Y_i$, and similarly for $Y$. This is the modular
weight of untwisted states for orbifold compactifications.
We also assume that the field $X$ has modular weight zero. The scalar potential 
comprises two terms: the usual scalar
potential of supergravity and the $D$ terms of the gauge
symmetry. Associated to $U(1)_X$ is the $D$ term
\begin{eqnarray}
V_{\rm D}  &=& \frac{g_X^2}{2}\biggl(K_XX-2K_YY+\sum_i q_iK_{Y_i}Y_i 
\nonumber \\
&+& \frac{g^2\hbox{Tr}(X)}{192\pi^2}m_{\rm Pl}^2\biggr)^2,
\end{eqnarray}
where $g_X$ is the $U(1)_X$ gauge coupling and $g$ the unified
string coupling. Modular invariance is compatible with the K\"ahler
potential
\begin{equation}
K=XX^*-\frac{3}{\kappa}\ln\biggl(Q+Q^*-\kappa\vert Y\vert^2-
\kappa\sum_i \vert Y_i\vert ^2\biggr).
\end{equation}
The $D$ term potential
vanishes altogether along a flat direction where the field $X$ acquires
a vacuum expectation value (vev) breaking the Abelian symmetry $U(1)_X$ at
\begin{equation}
\langle X \rangle =\sqrt
{\frac{-g^2 \hbox{Tr}(X)}{192\pi^2}}m_{\rm Pl},
\end{equation}
while the other fields vanish altogether. The value of the vev $\langle X\rangle $ is equal 
to a scale $\xi$ which is fixed in the heterotic string. This is not the case any more 
in the context of type I strings where $\xi$ is a moduli which is not fixed in perturbation 
theory. Expanding the superpotential in terms of Yukawa couplings, we
obtain  
\begin{equation}
W=\lambda (Q) X^2 Y +\cdots,
\end{equation}
where we have only taken into account the coupling such 
that $W_Y=\lambda (Q) \langle X\rangle ^2$ along the flat direction.
The function $\lambda (Q)$ is a modular form of weight $-2$. The scalar potential
along the flat direction is
\begin{equation}
V(Q)\propto \frac{\langle X \rangle ^4\lambda^2(1)}{12Q^2}.
\end{equation}
As already mentioned for the K\"ahler corrections, the kinetic 
term in such models is not standard. In the present case, the kinetic term for the 
moduli $Q$ is $[3/(4\kappa Q^2)](\partial
Q)^2$. Therefore, it is more convenient to redefine the fields such 
that the kinetic term becomes standard. This is achieved by means of 
the following expression: ${\rm d}\tilde{Q}=\sqrt 3/(\sqrt{2\kappa}Q){\rm d}Q$. The 
scalar potential then transforms into:
\begin{equation}
V(\tilde Q) \propto \frac
{\lambda^2(1)}{12}\langle X\rangle ^4e^{-2\sqrt{\frac{2\kappa}{3}}\tilde Q}.
\end{equation}
It has been shown in Refs. \cite{RP,FJ} that this potential also possesses 
remarkable tracking properties even if it has no inverse power factor of the 
quintessence field. However it suffers from phenomenological 
problems. First of all the slow decrease of the potential and the large value of the 
vev $\langle X \rangle $ imply that the quintessence field will have to be much larger than 
the Planck scale at the end of its evolution. This is a drawback as this would be in the 
string regime where the supergravity approximation is not valid anymore. In addition, 
one can show that the equation of state for this potential is such $\omega _Q=\omega _{\rm B}$ 
both in the radiation and matter dominated epochs \cite{FJ}. The scalar field follows 
exactly the behaviour of the background. As a consequence, the value of $\Omega _Q$ has 
been shown \cite{RP,FJ} to be limited to the relatively low value $\Omega _Q\approx 0.15$. This 
is not enough to reproduce the data which seem rather to indicate 
that $\Omega _Q\approx 0.7$. Nevertheless, the fact that the moduli field possesses 
a tracking behaviour is relevant as it could represent one of the components of the 
energy density of the universe. 

\subsection{An inverse power law  SUGRA model}
 
We now present a toy supergravity model with an inverse power law 
potential. To our opinion, this model is the most interesting 
one although we only present it as an existence proof of the 
quintessence property in supergravity.
\par
Let us use the same framework as in the previous section. We 
assume that the $U(1)_X$ Abelian symmetry is broken by the 
Fayet-Iliopoulos $D$ term. The superpotential is expanded as 
\begin{equation}
\label{superpotint}
W=\lambda X^2 Y+\dots ,
\end{equation}
where $\lambda$ is taken to be constant. 
This superpotential preserves the gauge 
symmetry of the model.
As claimed earlier we do find that $\langle W\rangle =0$ along the D-term flat 
directions. The main difference between the present model and the case of the moduli 
field of the heterotic string
is the different form of the K\"ahler potential. We choose the geometry of the moduli 
space to be singular at the origin, namely
\begin{equation}
K=XX^*+ \frac{(QQ^*)^p}{m_{\rm c}^{2p-2}}+\vert Y\vert^2 \frac{(QQ^*)^{n}}{m_{\rm c}^n}, 
\end{equation}
where $m_{\rm c}\approx 10^{15}\mbox {GeV}$ is the cut-off of the theory. It is 
reasonnable to choose it of the order of the unification scale. Higher order terms 
in $\vert Y\vert^2$ can also be included. This 
is the only relevant terms in the K\"ahler potential if one assumes the existence of a
modular symmetry. This modular symmetry is also important to prevent 
any K\"ahler correction. The scalar potential of this supergravity model is given by
\begin{equation}
V(Q)=\frac{\Lambda^{4+\alpha}}{Q^{\alpha}}e^{\frac{\kappa}{2}Q^2},
\end{equation}
where it is understood that $Q$ is now the canonically normalised 
field. The constant $\Lambda $ is given by 
$\Lambda^{4+\alpha}=2^{n/p}\lambda^2\langle X\rangle ^4 m_{\rm c}^{\alpha}$ 
where $\alpha=2n/p$. This potential has been studied in details in 
Ref. \cite{us}. There, it has been shown that, despite the appearence of positive powers 
of the field, the tracking properties are completely preserved. Using the 
constraint $\langle X\rangle \ge 10^2 \mbox{GeV}$ as the Abelian symetry has to 
be broken above the weak scale we find that the fine tuning 
problem can be overcome provided that $\alpha \ge 11$, see Ref. \cite{us}. This means 
that in order 
to obtain $\Omega _Q\approx 0.7$ today, one should fix $\Lambda $ to a value 
which is not far from the natural scales of high energy physics. For $\alpha =11$, actually 
we have $\Lambda \approx 10^{11}\mbox{GeV}$. The evolution of the 
energy density of the quintessence field is given in the following figure.

\begin{figure}
\begin{center}
\leavevmode
\hbox{%
\epsfxsize=8cm
\epsffile{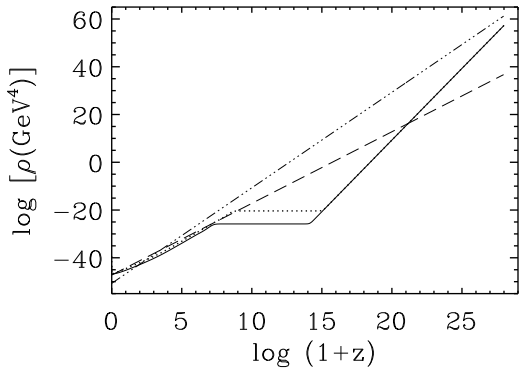}}
\end{center}
\caption{Energy density in the SUGRA model of quintessence for $\alpha =11$. The solid 
line is the energy density of quintessence whereas dashed-dotted and the dashed 
lines are the energy density of radiation and matter respectively. The dotted line 
is the energy density of quintessence for the potential 
$V(Q)=\Lambda ^{4+\alpha }Q^{-\alpha }$ for $\alpha =11$. The initial conditions are 
such that equipartition is realized just after inflation.}
\end{figure}

The evolution of the equation of state $\omega _Q$ for this model is displayed 
in the following figure. The evolution of the equation of state is almost 
unchanged during all the cosmic evolution. This is due to the fact that during 
this epoch $Q/m_{\rm Pl}$ is very small. As a consequence the exponential 
factor in the SUGRA potential plays no role and the SUGRA potential reduces 
to the usual potential. The situation changes at the end of the evolution. Since 
the field is on tracks one has $Q\approx m_{\rm Pl}$. This time the exponential 
factor in the potential plays a vital role and modifies the value of $\omega _Q$ 
today. 

\begin{figure}
\begin{center}
\leavevmode
\hbox{%
\epsfxsize=8cm
\epsffile{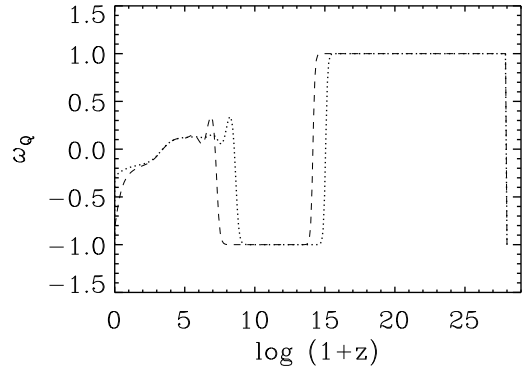}}
\end{center}
\caption{The dotted line represents the evolution of $\omega _Q$ for the 
potential $V(Q)=\Lambda ^{4+\alpha }Q^{-\alpha }$ with $\alpha =11$ whereas 
the dashed line represents the evolution of the equation of state 
for the same value of $\alpha $ and for the SUGRA model presented in this 
article.}
\end{figure}

We see that the value of the equation of state today, for $\alpha =11$, is given 
by \cite{us}:
\begin{equation}
\label{eqstate}
\omega _{Q}\approx -0.82.
\end{equation}
This is a remarkable property since in the context of usual tracking solutions 
it is not possible to obtain a number less than $\approx -0.7$ \cite{quint} whereas the 
observations seem to indicate that the value of the equation of state today 
is rather such that $-1 \le \omega _Q \le -0.8$. We would like to note that 
this nice property is almost independent of $\alpha $. This is illustrated 
on the following figure where the relation $\omega _Q-\alpha $ is displayed

\begin{figure}
\begin{center}
\leavevmode
\hbox{%
\epsfxsize=8cm
\epsffile{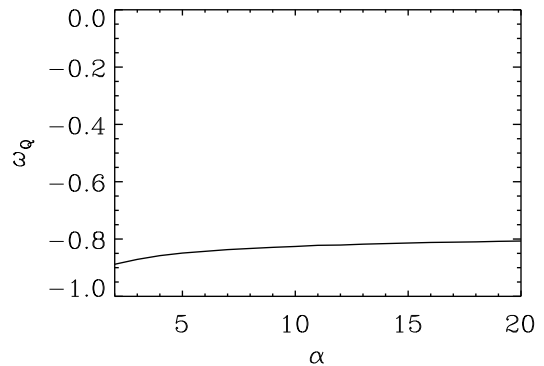}}
\end{center}
\caption{Equation of state $\omega _Q$ in function of $\alpha $ for the SUGRA 
model of quintessence proposed in this article.}
\end{figure}

The equation of state is almost independent of $\alpha $ because its value 
today is roughly speaking determined by the exponential term in the potential which 
is $\alpha $ independent. As a consequence, no fine tuning of $\alpha $ is required 
in order to obtain a reasonable value for $\omega _Q$. 
\par
The implications for structure formations of this model seem to 
be also very interesting and are currently under investigations \cite{BMR}.
\par
More details on this model can be found in Ref. \cite{us}.

\section{Conclusions}

We have studied the quintessence scenario recently proposed in Ref. \cite{track}. We 
have concentrated on models with an  inverse power law potential and the diverse corrections 
induced by its embedding into High Energy Physics models. More particularly we have 
focused on the quantum corrections in the non-supersymmetric setting,
the curvature and K\"ahler corrections in the supersymmetric case, and the effect of the
non-renormalisable interaction suppressed by the Planck mass in the supergravity context.
We have verified that quintessence is stable to the one-loop quantum corrections, 
preserving the existence of tracking solutions at small red-shift. We nevertheless 
argue that the solution to the fine-tuning problem
requires to consider the quintessence models within the framework of particle physics beyond 
the standard model. Emphasizing the supersymmetric scenario, the most likely candidate 
to describe the physics beyond the weak scale, we generalize the usual inverse power 
law potential of the $SU(N_{\rm c})$ super-QCD case to more
general supersymmetric gauge theories with Dynkin indices $\mu \le \mu_{\rm adj}$.
In supersymmetric models the quantum corrections to the superpotential vanish.
However there are two types of potentially dangerous corrections. Firstly the smallness 
of the effective mass
of the quintessence field implies that one must consider the small SUSY breaking induced 
by curvature effects.
We show that these effects are not a threat to the quintessence scenario. Secondly we 
investigate the corrections to the K\"ahler potential and show that they play only a 
role at the beginning of the evolution, decoupling at a scale of the order of the gaugino 
condensation scale, and are strong enough to jeopardize the
quintessence property. Far more relevant is the necessary inclusion of dangerous 
supergravity corrections
due to the presence of Planck mass suppressed interactions which induce negative 
contributions to the potential
at small red-shift. They destroy the quintessence property if the quintessence models 
are not treated within the realm of SUGRA.
\par
This is what we do in the final section considering models where the expectation 
value of the superpotential vanishes altogether. This prevents the appearance of the 
negative contributions to the energy density. We give two explicit models where this 
scenario is at work. One is based on the heterotic string at weak coupling, the role of 
the quintessence field being played by a moduli. The run-away behaviour of the moduli
is of the exponetial type limiting the possible contribution to the energy density 
to $\Omega _Q=0.15$ per moduli. In usual orbifold compactification this can amount 
to a $\Omega _Q=0.45$ contribution. Finally we present an existence proof
of an inverse power law potential in SUGRA by constructing a model based on a 
particular K\"ahler potential. This model is particularly promising as it provides a 
scenario for which the value of $\omega_Q$ is $-0.82$ for $\Omega _{\rm m}=0.3$. This is 
within one sigma of the recent experimental analyses \cite{Eft}.
\par
Let us emphasize that the models of quintessence within the framework of SUSY or SUGRA  
suffer from a problem raised in \cite{KL} concerning the necessary breaking of 
supersymmetry. Indeed there are two aspects to this problem, both linked to the types of 
corrections considered here. Let us first deal with the SUSY case. As advocated 
in \cite {KL} the present of a non-zero $F=M_{\rm S}^2$ term of the order $10^{10} \mbox{GeV}$ 
leads to an intolerably large value of the cosmological constant. This necessitates to 
invoke the prejudice already used in our paper that one does not know the mechanism which 
would force the cosmological constant to vanish. Quintessence scenarios do not aim at 
solving the quantum cosmological constant problem, see the conclusion of 
Ref. \cite{RP}. Another difficulty springs from the 
possible non-flatness of the K\"ahler potential and a coupling between the quintessence 
field $Q$ and the field responsible for the supersymmetry breaking leading to
a term given by:
\begin{equation}
{\rm \delta }V(Q)=k(Q)M_{\rm S}^4,
\end{equation}
where the series expansion of $k$ is a power series suppressed by the Planck scale. It 
gives a large contribution to the mass of $Q$ and the cosmological constant.
\par
Of course one should repeat the arguments within the SUGRA context. At tree level one can 
always fine-tune the minimum of the potential to be zero. The power law expansion of the 
potential might just give a correction to the quintessence behaviour which would only modify 
the end of the evolution of $Q$ while preserving quintessence. In the explicit model 
of section V.C, the SUSY breaking is induced by the $F$ terms of the 
dilaton and the moduli. The dilaton does not couple directly to the quintessence 
field Q in the K\"ahler potential because of the anomalous $U(1)_{\rm X}$ symmetry. This 
implies that our model is not affected by a dilatonic SUSY breaking. Assuming the 
existence of a modular symmetry, we find that the effect of the SUSY breaking 
by the moduli is to induce terms like $F^2/Q^{2p}$ in the scalar potential. More 
studies are required to evaluate the effect of this term, in particular 
its order of magnitude. However, the fact that this term is 
proportional to the inverse of the quintessence field suggests that it will cause 
no problems. This question will be adressed elsewhere \cite{us2}.
\par
In conclusion we would like to emphasize the explicit construction of particle physics models 
leading to the quintessence property is still to be further developped. It seems to us clear 
that the indications given in this paper show that the building of SUGRA models should turn out 
to be very fruitful.   

\section{Appendix I: The effective Action in Curved Space}

In the following we shall consider the case of curved spacetimes $R\ne 0$. 
Let us consider the following Lagrangian
\begin{eqnarray}
S &=& -\int {\rm d}^4 x\sqrt{-g}\biggr({\rm \partial}_{\mu} 
\varphi{\rm \partial}^{\mu} \varphi^* + 
i\bar\psi \bar \Sigma^{\mu} {\rm D}_{\mu} \psi \nonumber \\
&+& \frac{ m(\varphi)}{2}\psi\psi +
\frac{\bar m(\varphi)}{2} \bar\psi\bar\psi
+V(\varphi)\biggr),
\end{eqnarray}
where, for convenience, we have not written the spinorial indices. The mass and the 
potential are related to the superpotential $W$ as 
\begin{equation}
m(\varphi)=\frac {{\rm \partial}^2W}{{\rm \partial}\varphi^2},\quad V(\varphi)=\biggl\vert 
\frac{{\rm \partial}W}{{\rm \partial}\varphi}\biggr\vert ^2.
\end {equation}
This is the usual globally supersymmetric Lagrangian coupled to gravity. As supersymmetry is 
not preserved by the background curved space the Lagrangian receives quantum 
corrections. The effective potential is renormalised due to the
background geometry. Let us calculate this potential at the one loop level. 
\par
It is technically necessary  to perform a Wick rotation to obtain a Euclidean field 
theory on a curved
Riemannian manifold. This is globally feasible if one chooses the curved spacetime 
manifold to be a globally hyperbolic manifold with a Lorentzian signature. This 
implies that the time coordinate is globally defined allowing us to perform the
appropriate rotation $t \to it$. In order to have a good control on the 
Feynman path integral it is convenient to expand the bosonic field $\varphi$ 
around the classical solution $\varphi_{\rm c}$ of the Klein-Gordon 
equation $-\Delta \varphi_{\rm c}+({\rm \partial}V/{\rm \partial}\varphi_{\rm c}^*)=0$ 
where $\Delta $ is the curved Laplacian on the Riemannian manifold. Let us 
denote $\varphi=\varphi_{\rm c}+\tilde{\varphi }$ where $\tilde{\varphi }$ is a 
full-fledged quantum field. The effective potential does not involve derivatives 
of $\varphi_{\rm c}$, we will 
only retain non-derivative terms in the expansion of the effective action
\begin{equation}
e^{-S_{\rm eff}(\varphi_{\rm c})}=\int {\cal D} \tilde{\varphi}{\cal D}
\psi {\cal D}\bar\psi e^{-S},
\end{equation}
where $S_{\rm eff}$ is defined by the equation
\begin{equation}
S_{\rm eff}\equiv \int {\rm d}^4 x \sqrt{g} V_{\rm eff}(\varphi_{\rm c}).
\end{equation}
The effective potential is the sum of the classical part corrected by a 
quantum contribution
\begin{equation}
V_{\rm eff}(\varphi_{\rm c})= V(\varphi_{\rm c})+{\rm \delta}V(\varphi_{\rm c}).
\end{equation}
The correction term is due to the integration over the Weyl fermions and the bosonic 
fields. To leading order the effective action is
\begin{equation}
S_{\rm eff}(\varphi_{\rm c})=\mbox{tr}\ln (\Delta +\vert m\vert^2)-
\frac{1}{4}\mbox{tr}\ln (\Delta_{\rm F}),
\end{equation}
where $\Delta_{\rm F}$ is the operator 
$\Delta_{\rm F}={\rm D}_{\rm F}^{\dagger}{\rm D}_{\rm F}$ and 
\begin{equation}
{\rm D}_{\rm F}\equiv 
\left (
\begin {array}{cc}
 i{\rm D} & \bar m \\
m & i\bar {\rm D} \\
\end{array}
\right ),
\end{equation}
where ${\rm D}$ and $\bar{\rm D}$ are the Dirac operators acting on 
Weyl fermions of both chiralities. The previous expression can be computed 
using the $\zeta$ function regularisation where the determinant of an operator
is $\det A= \exp [-\zeta_{\rm A} ' (0)]$ and the $\zeta$ function 
is $\zeta (s) =\sum_n 1/\lambda_n^s$ as a function of the eigenvalues $\lambda_n$.
The $\zeta_{\rm A}$ function is related to the heat kernel solution of
\begin{equation}
AG(x,x',\sigma)=-\frac {\rm \partial}{{\rm \partial}\sigma}G(x,x',\sigma),
\end{equation}
with the boundary condition $G(x,x',0)=\delta (x-x')$ via the Mellin transform
\begin{equation}
\zeta_{\rm A}(s)=\frac{\mu^{2s}}{\Gamma (s)}\int_0^{\infty}{\rm d}\sigma \sigma^{s-1}
\int {\rm d}^4x \sqrt{g} G(x,x,\sigma).
\end{equation}
The scale $\mu$ is a renormalisation scale.
The determinant of the operator $A$  is essentially given by the asymptotic expansion 
of the heat kernel 
\begin{equation}
G(x,x,\sigma)\approx \frac {1}{(4\pi\sigma)^2}\biggl(1-\tau_1 \sigma+ O(\sigma ^2)\biggr),
\end{equation}
leading to
\begin{equation}
\det A=\exp\biggl(\int {\rm d}^4x \sqrt{g}\frac{\tau_1^2}{32\pi^2}
(\ln \frac {\tau_1}{\mu^2}-\frac{3}{2})\biggr).
\end{equation}
The asymptotic expansion of the Heat kernel of the operator $-\Delta +\vert m\vert ^2$ 
gives $\tau_1=-R/6+\vert m\vert ^2$ and for the Dirac operator
it yields $\tau_1^{\rm F}=\vert m\vert^2 +R/12$ where $R$ is the scalar 
curvature. The one loop effective potential is then
\begin{eqnarray}
{\rm \delta }V_{\rm eff} (Q,\eta ) &=& \frac {1}{32\pi^2 }
\biggl\{ \biggl[ \vert m\vert ^2 -\frac{R}{6}\biggr]^2
\biggl[\ln \biggl(\frac{\vert m\vert ^2-R/6}{\mu^ 2}\biggr)-\frac {3}{2}\biggr]
\nonumber \\
&-& \biggl[\vert m\vert ^2 +\frac{R}{12}\biggr]^2
\biggl[\ln \biggl(\frac{\vert m\vert ^2+R/12}{\mu ^2}\biggr)-\frac {3}{2}\biggr]\biggr\}.
\end{eqnarray}
Notice that the one loop correction vanishes if the curvature is zero.

\section{Appendix II: Supergravity Vacua}

In this appendix we shall be interested in finding whether a given
FLRW space time preserves supersymmetry. The FLRW background does
not break supersymmetry in an explicit manner when the gravitino
$\psi_{\mu}=0$ is invariant under supersymmetry transformations.
We have assumed that the quintessence field $Q$ is only coupled
to other field via the gravitational interactions. In a hidden
sector the supergravity can be broken with a non-vanishing 
mass $m_{3/2}$ leading to the condition
\begin{equation}
{\rm D}_{\mu }\epsilon+\frac{i}{2}m_{3/2}\sigma_{\mu}\bar \epsilon=0,
\end{equation}
where $\epsilon$ is a  Weyl spinor representing the supersymmetry variation.
Computing the commutator $[{\rm D}_{\mu}, {\rm D}_{\nu}]$ and using 
$R_{\mu\nu \rho\sigma}=(R/2)
(\eta_{\mu\rho}\eta_{\nu\sigma}-\eta_{\nu\rho}\eta_{\mu\sigma})$, we
find that the supersymmetry is generically broken apart from
two cases. If the gravitino mass vanishes, the supersymmetry is
fully preserved by FLRW spaces with vanishing curvature. When the
gravitino mass is not zero the supersymmetry is not broken if the
FLRW space is the anti-Desitter manifold with curvature
$R=-m_{3/2}^2$.

\acknowledgements

It is a pleasure to thank Martin Lemoine for useful exchanges and comments 
and for his invaluable help in the writing of the codes used in this paper.
We would like to thank Alain Riazuelo for useful discussions.


\begin{thebibliography}{99}

\bibitem{SNIa}
A.~G. Riess {\it et al.}, Astrophys.~J. {\bf 116}, 1009 (1998);
P.~M.~Garnavich {\it et al.}, Astrophys.~J. {\bf 509}, L74 (1998); 
S.~Perlmutter {\it et al.}, Astrophys.~J. {\bf 516}, (1999).

\bibitem{Teg}
M.~Tegmark, Astrophys.J. {\bf 514}, L69 (1999).

\bibitem{Sask}
C.~B.~Netterfield {\it et al.}, Astrophys. J. {\bf 474}, 47 (1997);
E.~Wollack {\it et al.}, Astrophys. J. {\bf 476}, 440 (1997);

\bibitem{PythonV}
K.~Coble {\it et al.}, {\tt astro-ph/9902195}.

\bibitem{Toco}
E.~Torbet {\it et al.}, {\tt astro-ph/9905100}.

\bibitem{velo}
I.~Zehavi and A.~Dekel, {\tt astro-ph/9904221}.

\bibitem{WCOS}
L.~Wang, R.~R.~Caldwell, J.~P.~Ostriker and P.~J.~Steinhardt, 
{\tt astro-ph/9901388}.

\bibitem{PTW}
S.~Perlmutter, M.~S.~Turner and M.~White, {\tt astro-ph/9901052}.

\bibitem{Eft}
G.~Efstathiou, {\tt astro-ph/9904356}.

\bibitem{Coleman}
S.~Coleman, Nucl. Phys. B {\bf 310}, 643 (1988);

\bibitem{quint}
R.~R.~Caldwell, R.~Dave and P.~J.~Steinhardt, 
Phys. Rev. Lett. {\bf 80}, 1582 (1998);

\bibitem{KL}
C.~Kolda and D.~H.~Lyth, {\tt hep-ph/9811375}.

\bibitem{track}
I.~Zlatev, L.~Wang and P.~J.~Steinhardt, {\tt astro-ph/9807002}; 
P.~J.~Steinhardt, L.~Wang and I.~Zlatev, {\tt astro-ph/9812313};

\bibitem{PB}
P.~Bin\'etruy, {\tt hep-ph/9810553}.

\bibitem{us}
P.~Brax and J.~Martin, {\tt astro-ph/9905040}.

\bibitem{Choi}
K.~Choi, {hep-ph/9902292}.

\bibitem{MPR}
A.~Masiero, M.~Pietroni and F.~Rosati, {\tt hep-ph/9905346}.

\bibitem{RP}
P.~J.~E~Peebles and B.~Ratra, Phys. Rev. D {\bf 37}, 3406  (1988).

\bibitem{CE}
S.~Coleman and E.~Weinberg, Phys. Rev. D {\bf 7}, 1888  (1973).

\bibitem{W}
S.~Weinberg, Phys. Rev. D {\bf 7}, 2887  (1973).

\bibitem{IIM}
J.~Iliopoulos, C.~Itzykson and A.~Martin, Rev. Mod. Phys.D {\bf 47}, 165  (1975).

\bibitem{skiba}
C.~Csaki, M.~Schmaltz and W.~Skiba, Phys. Rev. D {\bf 55}, 7840 (1997).

\bibitem{PBsusy}
P.~Brax, C.~Grojean and C.~A.~Savoy, {\tt hep-ph/9808345}.

\bibitem{G}
L.~P.~Grishchuk and Y.~V.~Sidorov, Phys. Rev. D. {\bf 42}, 3413 (1990).

\bibitem{WB}
J.~Wess and J.~Bagger, {\it Supersymmetry and Supergravity}, Princeton University Press.

\bibitem{Parker}
L.~Parker, Phys. Rev. D {\bf 3}, 346 (1971).

\bibitem{DSW}
M.~Dine, N.~Seiberg and E.~Witten, Nucl. Phys. B {\bf 289}, 589 (1988).

\bibitem{IL}
L.~E.~Ibanez and D.~Lust, Nucl. Phys. B {\bf 382}, 305 (1992).

\bibitem{FJ}
P.~G.~Ferreira and M.~Joyce, Phys. Rev. D {\bf 58}, 023503-1  (1998);
P.~G.~Ferreira and M.~Joyce, Phys. Rev. Lett. {\bf 79}, 4740 (1997);
C.~Wetterich, Nucl. Phys. B {\bf 302}, 668 (1988).

\bibitem{BMR}
P.~Brax, J.~Martin and A.~Riazuelo, in preparation.

\bibitem{us2}
P.~Brax, J.~Martin, in preparation.

\end{thebibliography}
\end{document}